\documentclass[12pt]{article}
\catcode`\@=11
\@addtoreset{equation}{section}

\global\arraycolsep=2pt
\usepackage{mathrsfs,amsbsy,amssymb,latexsym,amsfonts,amsmath,cite}
\usepackage{graphicx,color}
\usepackage{physics}
\usepackage{mathtools}
\usepackage{subfigure}
\usepackage{comment}

\newcommand{\wt}[1]{{\widetilde{#1}}}
\newcommand{\douti}{\Longleftrightarrow}

\newcommand{\vae}{\varepsilon}
\newcommand{\ifrac}[2]{{\left. #1 \middle/ #2 \right. }}

\newcommand{\aln}[1]{\begin{align}#1\end{align}}

\allowdisplaybreaks

\begin{document}
\renewcommand{\thefootnote}{\fnsymbol{footnote}}
\begin{flushright}
KUNS-2784
\end{flushright}
\vspace*{1.0cm}

\begin{center}
{\Large \bf Large N Analysis of $T\bar{T}$-deformation \vspace*{0.2cm}\\ 
and Unavoidable Negative-norm States} 
\vspace*{1.5cm}\\

{\large  Junichi Haruna\footnote{E-mail:~j.haruna@gauge.scphys.kyoto-u.ac.jp}, 
Takaaki Ishii\footnote{E-mail:~ishiitk@gauge.scphys.kyoto-u.ac.jp}, 
Hikaru Kawai\footnote{E-mail:~hkawai@gauge.scphys.kyoto-u.ac.jp},  \vspace*{0.2cm}\\
Katsuta Sakai\footnote{E-mail:~katsutas@gauge.scphys.kyoto-u.ac.jp}, 
and Kentaroh Yoshida\footnote{E-mail:~kyoshida@gauge.scphys.kyoto-u.ac.jp}} 
\end{center}

\vspace*{0.4cm}

\begin{center}
{\it Department of Physics, Kyoto University, Kyoto 606-8502, Japan} 
\vspace*{0.3cm}\\ 
\end{center}

\vspace{0.3cm}

\begin{abstract}\noindent
We study non-perturbative quantum aspects of $T\bar{T}$-deformation of 
a free $O(N)$ vector model by employing the large $N$ limit. It is shown that 
bound states of the original field appear and inevitably become negative-norm states. 
In particular, the bound states can be regarded as the states of the conformal mode 
in a gravitational theory, where the Liouville action is induced with the coefficient proportional 
to the minus of central charge. To make the theory positive-definite, 
some modification is required so as to preserve diffeomorphism invariance 
due to the Faddeev-Popov ghosts with a negative central charge.   
\end{abstract}

\setcounter{footnote}{0}
\setcounter{page}{0}
\thispagestyle{empty}

\newpage

\renewcommand\thefootnote{\arabic{footnote}}

\section{Introduction}
Recently, there has been much effort made to study $T\bar{T}$-deformation, 
which is an irrelevant perturbation by a composite operator, the determinant of 
the energy momentum tensor. Some peculiar characteristics of this operator 
were shown in 2004 by Sasha Zamolodchikov \cite{Zamolodchikov:2004ce}. 
In particular, the $T\bar{T}$-operator is well-defined as a composite operator 
only in {\it two} dimensions and the expectation value of the $T\bar{T}$-operator 
with non-degenerate stationary states exhibits the factorization 
for general 2d quantum field theories (QFTs). 

\medskip

In 2016, the $T\bar{T}$-deformation has been investigated in the context of 
2d integrable QFT \cite{Smirnov:2016lqw,Cavaglia:2016oda}. 
The quantum integrability is characterized by the novel factorization of S-matrix. 
Then the two-body S-matrix is determined by a bootstrap program up to the CDD factor 
\cite{Castillejo:1955ed}. Since the $T\bar{T}$-deformation is an irrelevant perturbation 
and cannot change the IR information like the mass pole structure in the S-matrix, 
the deformation can modify only the CDD factor. Thus, the S-matrix factorization is 
obviously preserved and in this sense, the $T\bar{T}$-deformation is an integrable deformation. 

%
%
%
%

\medskip 

In addition, by considering the effect of the CDD factor in the context of thermodynamic 
Bethe ansatz (TBA), a flow equation, called the $T\bar{T}$-flow equation, has been found 
\cite{Smirnov:2016lqw,Cavaglia:2016oda}, 
\aln{
\frac{d \mathcal{L}^{(\alpha)}}{d \alpha}= \det(T^{(\alpha)}_{\mu\nu})\,.  
\label{e:flow}}
Note here that the equality is shown for 2d integrable QFTs, but it is widely believed that 
the flow equation holds in more general setup\footnote{The $T\bar{T}$-deformation 
in closed form is presented in \cite{Bonelli:2018kik}. For a review, see for example 
\cite{Jiang:2019hxb}.}. Much investigation has been made in the $T\bar{T}$-deformation 
at the classical level. For example, by solving the flow equation, the $T\bar{T}$-deformation 
of free massless bosons is found to be the Nambu-Goto action in the static gauge, 
with $\alpha$ identified to the string tension \cite{Cavaglia:2016oda}. 
The $T\bar{T}$-deformation is also related to metric perturbation of 2d base space 
via a field-dependent coordinate transformation \cite{Dubovsky:2017cnj, Cardy:2018sdv, 
Conti:2018tca, Conti:2018jho, Ishii:2019uwk}. In this sense, the flow equation (\ref{e:flow}) 
at the classical level indicates an intimate connection to classical gravity. 

\medskip 

As another application, for 2d QFT on a cylinder, the $T\bar{T}$-flow equation can be rewritten 
into the inviscid Burgers equation  \cite{Smirnov:2016lqw,Cavaglia:2016oda}. In particular, 
in the case of conformal field theory (CFT), the usual CFT data can be used as the initial condition. 
Then by solving the Burgers equation, the energy spectrum of the $T\bar{T}$-deformed CFT 
can be exactly derived. The resulting spectrum has been confirmed by TBA 
\cite{Dubovsky1}. Here, an important observation is that while the energy spectrum 
has been well established by the $T\bar{T}$-flow equation or TBA, the associated states have not 
been studied so much. After all, from the field-theoretical point of view, it may be rather natural to suspect that the positivity should be broken for the high-energy states simply because the $T\bar{T}$-deformation is described by an irrelevant operator. 
In order to see the consistency of the system at the quantum level, 
it is significant to check the positivity of the states by non-perturbative method. 


\medskip

In this paper, we will tackle this issue by studying a $T\bar{T}$-deformed $O(N)$ vector model 
as a concrete example. In particular, we consider the large $N$ limit. Then the leading contribution 
of the path integral comes only from the configuration for a stationary point of the action. 
Hence all we have to analyze is just the classical solution. Remarkably, the degrees of freedom 
in the large $N$ limit represent the bound states of the original field. 
After evaluating their vacuum expectation values, we derive the kinetic terms for them. 
It should be emphasized that this large $N$ analysis can treat the full quantum correction. 

\medskip

We start from an infinitesimal $T\bar{T}$-deformation for simplicity. 
Section \ref{s:largeN} presents our crucial result that the induced kinetic term for the bound states 
has a wrong sign and leads to negative-norm states. Then, in Section \ref{s:grav}, 
its origin is elucidated by focusing upon the fact that the deformed theory may partially 
be seen as a theory coupled to gravity. From this section, we treat a more general deformation by an operator consisting of the energy-momentum tensor, which includes the original $T\bar{T}$-deformation. The negative-norm mode emerges in the general deformation as well and it is identified with 
the conformal mode in an ordinary gravitational theory, where the Liouville action 
is induced from matter loops. As long as the theory is approximated by a CFT 
in the high-energy limit, the coefficient of the Liouville action is proportional to 
the minus of central charge. This indicates that the conformal mode is positive-definite 
only when the total central charge is negative. This fact leads to the necessity to include 
something like the Faddeev-Popov (FP) ghosts, that is, some modification so as to make 
the theory diffeomorphism-invariant. In Section \ref{s:fin}, we generalize the preceding discussion on the infinitesimal deformation to the finite deformation case, and conclude that 
the deformed theory still contains a negative-norm state. We also make some discussion on the $1/N$ correction in Section \ref{s:1/N}. 
Finally, Section \ref{s:sum} is devoted to summary and discussion.

\section{Large $N$ analysis of $T\bar{T}$-deformation}
\label{s:largeN}

\subsection{Setup and gap equation}
\label{s:l}

In the following, we will consider an infinitesimal $T\bar{T}$-deformation of a free $O(N)$ vector 
model. The classical action is given by  
\aln{
S&=\int\!\! d^2x\, N\biggl[\frac{1}{2}(\partial_{\mu}\vec{\phi})^2-\frac{m_0^2}{2}\vec{\phi}^{\,2}+\alpha_0 \det(T_{\mu\nu})\biggr],
\label{e:act1}
\\
T_{\mu\nu}&\equiv\partial_{\mu}\vec{\phi}\cdot\partial_{\nu}\vec{\phi}
-\frac{\eta_{\mu\nu}}{2}\qty((\partial_{\rho}\vec{\phi})^2
-m_0^2\vec{\phi}^{\,2}).
}
Here $\vec{\phi} =\!\,^t(\phi_1,\cdots,\phi_N)$ is an $O(N)$ vector multiplet composed of $N$ real scalars 
satisfying $\vec{\phi}^{\,2} =\sum_{k=1}^N\phi_k^2$ and 
$T_{\mu\nu}$ is the energy-momentum tensor of the undeformed Lagrangian.
Also, $m_0^2$ and $\alpha_0$ are the bare mass and the bare deformation parameters, respectively. Note that $\alpha_0$ is normalized to be a {}'t Hooft coupling constant as $\alpha_0=N\alpha$. The flow equation with respect to $\alpha_0$ is the same form as Eq.(\ref{e:flow}), with the energy-momentum tensor replaced with the normalized one: 
$T^{(\alpha_0)}_{\mu\nu}=T^{(\alpha)}_{\mu\nu}/N$:
\aln{
\frac{d \mathcal{L}^{(\alpha_0)}}{d \alpha_0}= \det(T^{(\alpha_0)}_{\mu\nu})\,.  
\label{e:flow1}
}
In this paper, we will write all the quantity with the normalization associated with $\alpha_0$. 

\medskip

By introducing an auxiliary field $C^{\mu\nu}$, the classical action (\ref{e:act1}) can be rewritten as  
\aln{
\frac{S}{N}&=\int\!\! d^2x\, \left[\frac{1}{2}(\partial_{\mu}\vec{\phi})^2-\frac{m_0^2}{2}\vec{\phi}^{\,2}
-\frac{1}{2}T_{\mu\nu}C^{\mu\nu}+\frac{1}{8\alpha_0}\det(C^{\mu\nu})\right]\,, 
\label{e:act2}
}
where $C^{\mu\nu}$ is a symmetric tensor field. In the following, it is convenient to decompose $C^{\mu\nu}$ 
into the sum of the trace and traceless parts like 
\aln{
C^{\mu\nu}=\wt{C}^{\mu\nu}+\eta^{\mu\nu}C\,, 
}
where $\wt{C}^{\mu\nu}$ denotes the traceless part, and the trace of $C^{\mu\nu}$ is $C^\mu{}_\mu=2C$\,.  

\medskip

Our analysis is concerned with some loop divergence. Hence, in addition to the action (\ref{e:act2}), 
let us include the necessary counterterms in advance:
\aln{
\frac{S}{N}&=\int\!\! d^2x\,\left[\frac{1}{2}(\partial_{\mu}\vec{\phi})^2-\frac{m_0^2}{2}\vec{\phi}^{\,2}
-\frac{1}{2}T_{\mu\nu}C^{\mu\nu} +\frac{1}{8\alpha_0}\det(C^{\mu\nu}) \right.   \notag  \\[4pt] 
& \left. \hspace*{6cm} +\Lambda_0C+\beta_0(1+C)(\wt{C}^{\mu\nu})^2\right], 
\label{e:act3}}
where $\Lambda_0$ and $\beta_0$ are additional bare parameters that include divergent parts. 

\medskip

The partition function for the action (\ref{e:act3}) can be computed as 
\begin{align}
&\int\! \mathcal{D}\vec{\phi}\mathcal{D}C\mathcal{D}\wt{C}^{\mu\nu}
\exp\biggl[
iN\int d^2x\,\biggl(~
\frac{1}{2}\vec{\phi}\qty[-\partial^2-\qty(1+C)m_0^2-\partial_{\mu}\wt{C}^{\mu\nu}\partial_{\nu}]
\vec{\phi}\nonumber\\
&~~~~~~~~~~~~~~~~~~~~~~~~~~~~~~~~~~~~~~~~+\Lambda_0C+\beta_0(1+C)(\wt{C}^{\mu\nu})^2+\frac{1}{8\alpha_0}(\wt{C}^{\mu\nu})^2
-\frac{1}{4\alpha_0}C^2
\biggr)\biggr] \notag
\\ 
\propto & \int\! \mathcal{D}C\mathcal{D}\wt{C}^{\mu\nu}
\exp\biggl[
-\frac{N}{2}\tr\log\qty[\partial^2+\qty(1+C)m_0^2+\partial_{\mu}\wt{C}^{\mu\nu}\partial_{\nu}-i\vae]\nonumber\\
&~~~~~~~~~~~~~~~~~~~~~~~~~~~~+iN\int d^2x\,\qty(\Lambda_0C+\beta_0(1+C)(\wt{C}^{\mu\nu})^2+\frac{1}{8\alpha_0}(\wt{C}^{\mu\nu})^2-\frac{1}{4\alpha_0}C^2)
\biggr]\,, \notag
\end{align}
where $\partial^2\equiv\eta^{\mu\nu}\partial_\mu\partial_\nu$. 
Thus, the effective action for $C$ and $\wt{C}_{\mu\nu}$ is given by 
\begin{align}
\frac{\Gamma}{N}
&=\frac{i}{2}\tr\log\qty[\partial^2+\qty(1+C)m_0^2-i\vae+\partial_{\mu}\wt{C}^{\mu\nu}\partial_{\nu}]\nonumber\\
&~~~+\int\!\! d^2x\,\qty(\Lambda_0C+\beta_0(1+C)(\wt{C}^{\mu\nu})^2+\frac{1}{8\alpha_0}(\wt{C}^{\mu\nu})^2-\frac{1}{4\alpha_0}C^2).
\label{e:effaction}
\end{align}

\medskip

In the large $N$ limit, a single configuration for which the fields take the stationary point of the effective action  
can contribute to the integration over $C$ and $\wt{C}_{\mu\nu}$\,.  
The stationary condition of the action determines the vacuum expectation values (VEVs) of the auxiliary fields.
Intuitively, the VEVs represent the ones of the square of the original variables like 
\begin{align}
\vev{C}\sim\vev{\vec{\phi}^{\,2}}\,, \qquad 
\vev{\wt{C}^{\mu\nu}} \sim\vev{(\partial^\mu\vec{\phi}\cdot\partial^\nu\vec{\phi})_\mathrm{traceless}}\,. 
\end{align} 
Note here that $\vev{\wt{C}^{\mu\nu}}=0$ because $\wt{C}^{\mu\nu}$ is traceless and the vacuum is invariant 
under the translations and the Lorentz transformations.
This value satisfies the condition of the stationary action. 

\medskip

On the other hand, the condition for $\expval{C}$ is given by the gap equation (or the quantum equation of motion): 
\begin{align}
\fdv{S_{\mathrm{eff}}}{C}\eval_{C=\expval{C},\wt{C}^{\mu\nu}=0}&=0\\
\douti~~\frac{im_0^2}{2}\tr\qty[\frac{1}{\partial^2+\qty(1+\expval{C})m_0^2-i\vae}]&+\Lambda_0-\frac{1}{2\alpha_0}\expval{C}=0\,.
\label{e:gap}
\end{align}

\medskip

Here we regulate the first term on the left-hand side of Eq.\,(\ref{e:gap}) by Pauli-Villars (PV) regulators.
It turns out two regulators are necessary because quadratic and logarithmic divergences appear in the following calculation.  
The Lagrangian for the PV regulators is given by
\begin{align}
\frac{\mathcal{L}_{\rm PV}}{N}=\sum_{i=1,2}\sum_{r=1}^{d_i}
\frac{1}{2}\vec{\psi}_{ir}\qty[-\partial^2-\qty(1+C)M_i^2
-\partial_{\mu}\wt{C}^{\mu\nu}\partial_{\nu}]\vec{\psi}_{ir}\,, 
\end{align}
where $\vec{\psi}_{ir}$ are $O(N)$ multiplets with masses $M_i$.
Here we assume that for every $i$, $\{\vec{\psi}_{ir}\}_{r=1}^{d_i}$ are bosonic and each $d_i$ 
can be either positive or negative fraction. The divergence in the loop integrals can be removed 
by tuning the parameters as
\begin{align}
1+\sum_i d_i&=0\,, \qquad 
m_0^2+\sum_i d_iM_i^2=0\,.
\end{align}
Since the total energy-momentum tensor is the sum of the energy-momentum tensor for each field,
the regularized trace term in (\ref{e:gap}) is the sum of the contributions from the original field and the PV regulators.

\medskip

By adopting the PV regularization, the regularized trace term in (\ref{e:gap}) 
is evaluated as
\begin{align}
\notag
&\frac{im_0^2}{2}\tr\qty[\frac{1}{\partial^2+(1+\expval{C})m_0^2-i\vae}]
\\&\rightarrow\frac{i}{2}\int\!\!\frac{d^2k}{(2\pi)^2}\, \qty(\frac{m_0^2}{-k^2
+(1+\expval{C})m_0^2-i\vae}+\sum_i \frac{d_iM_i^2}{-k^2+(1+\expval{C})M_i^2-i\vae}) \notag
\\&=\frac{1}{8\pi}\qty[m_0^2\log(\frac{m_0^2}{M_1^2})-\frac{r\log(r)}{1-r}M_1^2
+\frac{r\log(r)}{1-r}m_0^2],
\end{align}
where $r\equiv\ifrac{M_2^2}{M_1^2}$. 
As a result, the gap equation is rewritten as 
\begin{align}
&\Lambda_0+\frac{1}{8\pi}\qty[m_0^2\log(\frac{m_0^2}{M_1^2})-\frac{r\log(r)}{1-r}M_1^2+\frac{r\log(r)}{1-r}m_0^2]=\frac{1}{2\alpha_0}\expval{C}.
\label{e:gapur}
\end{align}
This equation contains the quadratic and logarithmic divergent parts as $M_1\rightarrow \infty$\,. 
But these are controlled by renormalizing $\Lambda_0$, not $\alpha_0$! 
\!\!\footnote{
Although the renormalization of $\alpha_0$ is discussed in \cite{Rosenhaus:2019utc}, our interpretation here is different. 
}\,\,\,
This is the reason why the $\Lambda_0 C$ term was added to the action (\ref{e:act2}) in (\ref{e:act3}) in advance. 

\medskip

The renormalized parameter $\Lambda$ can be defined as 
\aln{
\Lambda&\equiv\Lambda_0+\frac{1}{8\pi}\qty[m_0^2\log(\frac{m_0^2}{M_1^2})-\frac{r\log(r)}{1-r}M_1^2+\frac{r\log(r)}{1-r}m_0^2]\,.
\label{e:gapur2}
}
Thus $\expval{C}$ is proportional to $\Lambda$ like 
\aln{
\expval{C}=2\alpha_0\Lambda\,.
}

\subsection{Two-point function and propagating degrees of freedom}
\label{s:subl}
Let us next evaluate a two-point function of ${C}^{\mu\nu}$ to see the propagating degrees of freedom. 

\medskip

The fluctuation of $C$ around the VEV can be described by $C'$ as  
\begin{align}
C=\expval{C}+C'\,.
\end{align}
It is later convenient to define a dressed mass $m$ as 
\aln{
m^2\equiv(1+\expval{C})m_0^2\,.
}
Then the action (\ref{e:effaction}) can be expanded as, 
up to second order in $C'$ and $\wt{C}^{\mu\nu}$, 
\begin{align}
\frac{\Gamma}{N}=-\frac{i}{4}&\tr\qty[\frac{1}{-\partial^2-m^2+i\varepsilon}
\qty(-m_0^2C'-\partial_{\mu}\wt{C}^{\mu\nu}\partial_{\nu})]^2+\frac{1}{\alpha_0}\int\!\! d^2x
\biggl[\frac{1}{8}(\wt{C}^{\mu\nu})^2-\frac{1}{4}C^{\prime2}\biggr]\nonumber \\[4pt]
&+(1+\expval{C})\beta_0\int\!\! d^2x\, (\wt{C}^{\mu\nu})^2+O((C',\wt{C}^{\mu\nu})^3).
\label{e:setwo}
\end{align}
The zeroth-order term has been dropped because it is just a constant term. 
The first-order term vanishes due to the gap equation.

\medskip

In a generic regularization scheme, the two-point function of $\wt{C}^{\mu\nu}$ is quadratically divergent. Expanding them by the external momentum, we will have some kinetic terms with logarithmically divergent coefficients. We can cancel them by introducing the corresponding counterterm to Eq.\,(\ref{e:act3}), which takes the following form:
\aln{
\delta(\partial_\lambda \wt{C}^{\mu\nu})^2,
\label{e:kin1}
}
where $\delta$ is a logarithmically divergent coefficient. 
\!\!\!\footnote{
In a general dimension, there are two kinds of kinetic terms for $\wt{C}^{\mu\nu}$: $(\partial_\lambda \wt{C}^{\mu\nu})^2$ and $(\partial_\mu\wt{C}^{\mu\nu})^2$. In 2d spacetime, however, these terms are not independent due to the following identity:
\aln{
\int\!\frac{d^2p}{(2\pi)^2}\,F(p)\wt{C}_\mu^{~\lambda}(p)\wt{C}_\lambda^{~\nu}(-p)=\int\!\frac{d^2p}{(2\pi)^2}\,F(p)\frac{1}{2}\delta_\mu^{~\nu}\wt{C}_\lambda^{~\rho}(p)\wt{C}_\rho^{~\lambda}(-p),
}
where $F(p)$ is an even function of $p_\mu$. 
}\,\,\,
However, this term leads to a nonlocal form of $T_{\mu\nu}$, and hence of $\vec{\phi}$, when we eliminate $C^{\mu\nu}$ from Eq.\,(\ref{e:act3}) with Eq.\,(\ref{e:kin1}).

Even if we allow such a counterterm, the corresponding finite part in the effective action, $b(\partial_\lambda \wt{C}^{\mu\nu})^2$, causes another problem, namely violation of positivity. Regardless of the sign of $b$, this term necessarily gives rise to negative-norm states. In fact, $(\partial_\lambda\wt{C}^{\mu\nu})^2$ includes kinetic terms for the components of $\wt{C}^{\mu\nu}$, which have relatively opposite signs:
\aln{
(\partial_\lambda\wt{C}^{\mu\nu})^2=(\partial_\lambda\wt{C}^{00})^2
-(\partial_\lambda\wt{C}^{0i})^2+(\partial_\lambda\wt{C}^{ij})^2.
}
Therefore, at least one mode is of negative norm whether we tune the sign of $b$ positive or negative. 
Thus the coefficient $b$ must be tuned to be zero in order to avoid the violation of positivity. 

\medskip

Even if we set $b=0$, the quadratic part of the effective action contains a wrong-sign  kinetic term for a specific component of $C^{\mu\nu}$. In fact, by PV regularization we obtain
\begin{align}
\frac{\Gamma|_\text{quad}}{N}=\int\!\! \frac{d^2p}{(2\pi)^2}\,
\biggl[&
\frac{1}{2}g(p)\qty(\frac{1}{1+\expval{C}}C'(-p)-\frac{p_\mu p_\nu}{p^2}\wt{C}^{\mu\nu}(-p))
\qty(\frac{1}{1+\expval{C}}C'(p)-\frac{p_\mu p_\nu}{p^2}\wt{C}^{\mu\nu}(p))\nonumber\\
&~~~~~+\qty(\frac{1}{8\alpha_0}+(1+\expval{C})\beta)\wt{C}^{\mu\nu}(-p)\wt{C}_{\mu\nu}(p)-
\frac{1}{4\alpha_0}C'(-p)C'(p)
\biggr],
\label{e:ser}
\end{align}
where the scalar function $g(p)$ and the renormalized parameter $\beta$ are defined respectively as 
\aln{
g(p)&\equiv -\frac{m^2}{8\pi }-\frac{p^2}{48 \pi }+\frac{m^4 }{2\pi  p \sqrt{4 m^2-p^2}}\tan ^{-1}
\left(\frac{p}{\sqrt{4 m^2-p^2}}\right)\,,\\
\beta&\equiv \beta_0-\frac{1}{32\pi}\biggl[m_0^2\log(\frac{m_0^2}{M_1^2})-\frac{r\log(r)}{1-r}M_1^2+\frac{r\log(r)}{1-r}m_0^2\biggr].
\label{e:beta}
}
For the detail of the calculation, see Appendix \ref{s:cal}. Note here that the first term in Eq.\,(\ref{e:ser}) is nonlocal in 
$\wt{C}^{\mu\nu}$ because of the factor $1/p^2$. Therefore its overall coefficient, 1/2, cannot be changed by any finite renormalization, and hence is determined independently of the regularization scheme.

\medskip

Let us read off the propagating degrees of freedom from Eq.\,(\ref{e:ser}) in the high-energy limit, $|p|\to \infty$. In this limit, $g\to-p^2/48\pi$\,.
\!\!\!\footnote{
While $g(p^2)$ has a branch point $p^2=4m_0^2$ in the complex $p^2$ plane, the behavior 
$g(p^2\to\infty)\sim -p^2/48\pi$ is independent of branches.
}\,\,\,
Then one can see that only one degree of freedom in $C^{\mu\nu}$,
\aln{
\chi\equiv\frac{1}{1+\expval{C}}C'-\frac{p_\mu p_\nu}{p^2}\wt{C}^{\mu\nu},
\label{e:chi2}
}
propagates, and the sign of its kinetic term becomes 
asymptotically {\it negative}.

If we employ a generic regularization scheme, the kinetic term for $C'$ is modified while that for $(p_\mu p_\nu/p^2)\wt{C}^{\mu\nu}$ is not. 
Therefore, the kinetic term in the high-energy limit is written as
\aln{
\frac{\Gamma|_\text{quad}}{N}~\rightarrow~\frac{1}{2}\int \frac{d^2p}{(2\pi)^2} ~p^2\,\mqty(\chi(p)&C'(p))\mqty(-\frac{1}{48\pi}&t\\t&u)\mqty(\chi(-p)\\C'(-p)),
\label{e:mix1} 
}
An important point is that the eigenvalues of the matrix in Eq.\,(\ref{e:mix1}) always include at least one negative value, regardless of the values of $t$ and $u$. It can be easily seen from the behavior of Eq.\,(\ref{e:mix1}) for large $\chi$. 

\medskip

In this manner, the emergence of negative-norm state from $\chi$ is inevitable even if we set $b=0$ by renormalization. This is a non-perturbative result for the $T\bar{T}$-deformed theory at the quantum level. 
Recall that $C^{\mu\nu}$ stands for composite operators made of $\vec{\phi}$ at the tree level. 
This means that bound states of two $\vec{\phi}$ 
are contained in the $T\bar{T}$-deformed theory as negative-norm states. 

\medskip

It is worth mentioning that the PV regularization or the dimensional regularization automatically realizes the tuning $b=0$. 
That is, the kinetic term of $\wt{C}^{\mu\nu}$ is identically zero 
with these regularizations. We further note that the second term in Eq.\,(\ref{e:beta}) is just the minus quarter of that in Eq.\,(\ref{e:gapur2}). It means that we can tune the bare parameter as $\beta_0=-\Lambda_0/4$, although the renormalizations of $\Lambda$ and $\beta$ are independent in the first place. The underlying structure of these facts will be demystified in the next section.

\section{Relation to gravity}
\label{s:grav}

In this section, we shall reveal why the wrong-sign kinetic term is induced 
as a radiative correction. We also answer some questions which one might 
have noticed in the previous section and Appendix. Why is the divergent part in $\beta_0$ just a quarter of 
that in $-\Lambda_0$\,? Why do we have no local kinetic term for ${C}^{\mu\nu}$ induced under 
the PV regularization or the dimensional regularization? Why does only one degree of freedom propagate among $C^{\mu\nu}$? The answers to these questions will be provided by comparing 
the deformed theory and a theory coupled to 2d gravity.  

\medskip


In the following, let us consider a generalization of $T\bar{T}$-deformation 
described by a modified flow equation, 
\aln{
\frac{d \mathcal{L}^{(\alpha_0)}}{d \alpha_0}= f(T^{(\alpha_0)}_{\mu\nu})\,, 
\label{e:flow2}
}
where $f(T^{(\alpha_0)}_{\mu\nu})$ is an arbitrary function of the energy-momentum tensor 
$T^{(\alpha_0)}_{\mu\nu}$ for the deformed Lagrangian $\mathcal{L}^{(\alpha_0)}$. 
The original 
$T\bar{T}$-deformation is contained as a special case. Note here that the quantum integrability is 
not ensured by the modified flow equation (\ref{e:flow2}) in general. 

\medskip 

According to this modification, 
the action for an infinitesimal deformation is given by
\aln{
S&=\int\!\! d^2x\, N\biggl[\frac{1}{2}(\partial_{\mu}\vec{\phi})^2-\frac{m_0^2}{2}
\vec{\phi}^{\,2}+\alpha_0 f(T_{\mu\nu})\biggr]\,.
\label{e:act5}
}
Note here that Eq.\,(\ref{e:act5}) 
is equivalent to the following action: 
\aln{
\frac{S}{N}&=\int\!\! d^2x\,\left[\frac{1}{2}(\partial_{\mu}\vec{\phi})^2-\frac{m_0^2}{2}\vec{\phi}^{\,2}
-\frac{1}{2}T_{\mu\nu}C^{\mu\nu}+h\left(C^{\mu\nu}; \alpha_0\right)\right], 
\label{e:act6}
} 
where the function $h(C^{\mu\nu};\alpha_0)$ is determined so that Eq.\,(\ref{e:act5}) is reproduced 
after removing $C^{\mu\nu}$ with the use of the equation of motion for $C^{\mu\nu}$.

\medskip

In order to compare the theory described by (\ref{e:act6}) with the $O(N)$ vector model 
coupled to gravity, let us rewrite the classical action (\ref{e:act6}) as
\aln{
\frac{S}{N}&=\!\!\int\!\! d^2x\,\sqrt{-g'}\,\left[-\det(\eta^{ab}-\wt{C}^{ab})\right]^{\!\!\frac{1}{2}}\qty[\frac{1}{2}g^{\prime\mu\nu}\partial_\mu\vec{\phi}\cdot\partial_\nu\vec{\phi}-\frac{m_0^2}{2}\vec{\phi}^{\,2}]
+\!\!\int\!\! d^2x\,h\nonumber\\[4pt] 
&=\!\!\int\!\! d^2x\,\sqrt{-g'}\,\qty[\frac{1}{2}g^{\prime\mu\nu}\partial_\mu\vec{\phi}\cdot\partial_\nu\vec{\phi}-\frac{m_0^2}{2}\vec{\phi}^{\,2}] \nonumber \\ 
& \quad +\int\!\! d^2x\,\sqrt{-g'}\,\Psi\qty[\frac{1}{2}g^{\prime\mu\nu}\partial_\mu\vec{\phi}\cdot\partial_\nu\vec{\phi}-\frac{m_0^2}{2}\vec{\phi}^{\,2}] +\!\!\int\!\! d^2x\,h,
\label{e:act7}
}
where we have introduced new quantities: 
\aln{
g^{\prime\mu\nu}& \equiv \frac{\eta^{\mu\nu}-\wt{C}^{\mu\nu}}{1+ C}, 
\qquad \sqrt{-g'}=\left[-\det(g^{\prime\mu\nu})\right]^{-\frac{1}{2}},
\label{e:met1}
\\
\Psi&\equiv\left[-\det(\eta^{ab}-\wt{C}^{ab})\right]^{\!\frac{1}{2}}-1 =-\frac{1}{4}(\wt{C}^{\mu\nu})^2+O((\wt{C}^{\mu\nu})^3).
}
It should be remarked that $C$ and $\wt{C}^{\mu\nu}$ are not infinitesimal fluctuations 
but parametrize the metric as $g^{\prime\mu\nu}$, and that Eq.\,(\ref{e:act7}) is exactly the same action as 
Eq.\,(\ref{e:act6}) for finite $C^{\mu\nu}$.
This action is not invariant under diffeomorphism. 
In fact, the diffeomorphism invariance is broken by the determinant factor 
$\mathrm{det}(\eta-\wt{C})^{1/2}$ (or $\Psi$ equivalently), and by the last term unless $h=\sqrt{-g'}$. 
Recall that diffeomorphism invariance plays a crucial role in making a gravitational theory positive-definite.  
Then it looks quite in question whether the present theory is positive-definite or not. 
Indeed, this lack of diffeomorphism invariance is exactly the source for the positivity-violating kinetic term in Eq.\,(\ref{e:ser}). 

\medskip

Let us re-derive the previous result by considering integration of $\vec{\phi}$ 
in the action written as Eq.\,(\ref{e:act7}). Note that the radiative correction is the same as 
that of the theory coupled to the ordinary gravity, except for the $\Psi$ contribution. 
However, $\Psi$ has no influence on the two-point function. Since $\Psi$ is $O((\wt{C}^{\mu\nu})^2)$, 
it could affect the mass term of the two-point function through the tadpole diagram for it. 
The contribution from the diagram eventually vanishes because the interaction vertex for $\Psi$ 
and $\vec{\phi}$ is just the inverse of the propagator for $\vec{\phi}$: 
\aln{
\text{(tadpole for $\Psi$)} &= \int\!\!\frac{d^2p}{(2\pi)^2}\frac{i}{p^2-m_0^2+i\vae}(-i)(p^2-m_0^2)\nonumber\\
&=\int\!\!\frac{d^2p}{(2\pi)^2}1~~\xrightarrow{\text{regularized}}~0.
}
Therefore, as long as we investigate the two-point function, we obtain exactly the same result as in the ordinary gravitational theory. 

\medskip 

There are three essential points to be explained: 1) the necessity of the cosmological constant for renormalization, 2) the induction of the kinetic terms as the Einstein-Hilbert (EH) action, and 3) the induction of the Liouville action in a CFT.

\subsubsection*{1) the necessity of the cosmological constant for renormalization}

Let us consider the following action instead of Eq.\,(\ref{e:act7}):
\aln{
\frac{S}{N}&=\!\!\int d^2x\sqrt{-g'}\qty[\frac{1}{2}g^{\prime\mu\nu}\partial_\mu\vec{\phi}\cdot\partial_\nu\vec{\phi}-\frac{m_0^2}{2}\vec{\phi}^{\,2}+\Lambda_0]
+\int d^2x\sqrt{-g'}\Psi\qty[\frac{1}{2}g^{\prime\mu\nu}\partial_\mu\vec{\phi}\cdot\partial_\nu\vec{\phi}-\frac{m_0^2}{2}\vec{\phi}^{\,2}]
\nonumber\\
&~~~~~~~~~~~~~~~~~~~~~~~~~~~~~~~~~~~~~~~~~~~~~~~~~~~~~~~~~+\!\!\int d^2x\,h.
\label{e:act8}
}
By expanding the cosmological constant term up to the second order in $C^{\mu\nu}$, we obtain
\aln{
\int d^2x\sqrt{-g'}\Lambda_0=\text{(const.)}+\Lambda_0C-\frac{1}{4}\Lambda_0(\wt{C}^{\mu\nu})^2+O((C,\wt{C}^{\mu\nu})^3).
}
This is why we had a choice to set the bare parameter as $\beta_0=-\Lambda_0/4$ in the previous section. 

\subsubsection*{2) the induction of the kinetic terms as the Einstein-Hilbert (EH) action}

Recall that in a gravitational theory, matter loops induce the kinetic term for the metric in the form of the EH action, as long as we adopt a regularization preserving the diffeomorphism invariance. 
Since the PV and dimensional regularizations preserve the diffeomorphism invariance,  
the kinetic terms for $C^{\mu\nu}$ in Eq.\,(\ref{e:ser}) might be obtained as the linearized EH action. 

\medskip

However, the combination of the kinetic terms identically vanishes because the EH action is topological 
in two dimensions. This fact corresponds to the absence of local kinetic terms such as Eq.\,(\ref{e:kin1}) in Eq.\,(\ref{e:ser}). 

\subsubsection*{3) the induction of the Liouville action in a CFT}

The third point is the most critical in our analysis. Recall that in the case of 2d CFT coupled to gravity, the integration of a matter field induces the Liouville action: 
\aln{
S_\text{L}=\frac{1}{2}\int\!\! d^2xd^2y~\sqrt{-g(x)}R(x)D(x-y)\sqrt{-g(y)}R(y), 
}
where $D(x-y)$ is the inverse operator of d'Alembertian. The crucial point is that the coefficient of $S_\text{L}$ in the effective action is proportional to the minus of the central charge for the integrated field. If we take the loop contribution only from the ordinary matter field, the coefficient is of wrong sign for $S_\text{L}$, because it represents the kinetic term of the conformal mode of the metric. This discussion is applicable even to a general 2d field theory as long as it is approximated by a CFT in the high-energy limit. 

\medskip

Indeed, the quadratic part of the effective action (Eq.\,(\ref{e:ser})) approaches that of $S_\text{L}$ 
in the high-energy limit $p^2\to\infty$: 
\aln{
\Gamma|_\text{quad}&\rightarrow-\int\frac{d^2p}{(2\pi)^2}\frac{N}{96\pi}p^2\chi(-p)\chi(p)\\
&=\frac{-N}{48\pi}S_\text{L}\Big|_{g^{\mu\nu}=g^{\prime\mu\nu},\text{  quad}}
}
with
\aln{
g^{\prime\mu\nu}=\frac{\eta^{\mu\nu}-\wt{C}^{\mu\nu}}{1+\expval{C}+C'}
}
and $\chi$ defined by Eq.\,(\ref{e:chi2}). Note that each component of $\vec{\phi}$ approaches to the conformal matter with a central charge 
$c=+1$ in the high-energy limit. The coefficient of $S_\text{L}$ consistently appears with the total matter contribution to the central charge $c_{\rm tot} = N$. Now, the meaning of the only propagating mode 
is obvious; it is the conformal mode. Consider a reparametrization of $C'$ and $\wt{C}^{\mu\nu}$ 
according to the diffeomorphism and conformal parts:
\aln{
g'_{\mu\nu}
&=(1+\expval{C}+C')(\eta_{\mu\nu}+\wt{C}_{\mu\nu}+\wt{C}_\mu^{~\lambda}\wt{C}_{\lambda\nu}+\cdots)\nonumber\\
&=(1+\expval{C})(\eta_{\mu\nu}+\eta_{\mu\nu}\chi+\partial_\mu\xi_\nu+\partial_\nu\xi_\mu).
\label{e:metrep}
}
Here, $\chi$ is identical to that defined by Eq.\,(\ref{e:chi2}), which is the only propagating mode. 
Therefore, we conclude that the general deformation (Eq.\,(\ref{e:act5})), even equipped 
with the necessary counter terms (Eq.\,(\ref{e:act7})), violates the positivity at the quantum level. 

\medskip

The only possible way to flip the sign of the Liouville action is to introduce ghosts. 
The healthy ghosts that do not cause any pathology in the context of the ordinary field theory 
are nothing but the FP ghosts. This discussion 
compels us to recover the diffeomorphism-invariance in the theory and to take account of  
the corresponding FP ghosts. 

\medskip

If the action could be equipped with diffeomorphism-invariance,  
the diffeomorphism part of the field can be gauged away and then the effective action 
in the large $N$ limit would take the following form:
\aln{
\Gamma=\left(\frac{26-N}{48\pi}\,S_\text{L}+\int d^2x\sqrt{-g}\,\Lambda\right)\bigg|_{g_{\mu\nu}=\eta_{\mu\nu}(1+\chi)}+\int \frac{d^2p}{(2\pi)^2}\biggl[p^2O\left(\frac{m_0^2}{p^2}\right)+O(\chi^3)\biggr],
\label{e:SL3}
}
where the number 26 has come from the central charge of the FP ghosts. It means that the theory will be positive-definite with a condition $N<26$. This observation might be related to \cite{Callebaut:2019omt}, where a $T\bar{T}$ deformed CFT is proposed to be a gauge-fixed  noncritical string, and the Virasoro condition is taken into account to make the total central charge vanish. The diffeomorphism in the $T\bar{T}$-deformation is discussed by focusing on its equivalence to the massive gravity theory \cite{Tolley:2019nmm}.

\section{Unavoidable negative-norm states in finite deformation}
\label{s:fin}

So far, we have discussed the positivity of the infinitesimally $T\bar{T}$-deformed $O(N)$ 
vector model. In this section, we will show that the negative norm state exists 
even in the case of finite deformation.  

\medskip

To get a solution to Eq.\,(\ref{e:flow2}), it is useful to rewrite the Lagrangian with auxiliary fields 
like Eq.\,(\ref{e:act6}). For this purpose, let us introduce new quantities: 
\aln{
s_{\mu\nu}&\equiv\frac{1}{2}\partial_\mu\vec{\phi}\cdot\partial_\nu\vec{\phi}\,,  \qquad M\equiv\frac{m_0^2}{2}\vec{\phi}^{\,2}\,.
}
Then the undeformed Lagrangian is expressed 
only by these quantities as
\aln{
\frac{\mathcal{L}^{(0)}}{N}
&=g^{\mu\nu}s_{\mu\nu}-M\Big|_{g_{\mu\nu}=\eta_{\mu\nu}}.
\label{e:ini1}
}
Since the energy-momentum tensor is defined as the functional derivative of 
the covariantized action by $g_{\mu\nu}$\,, the solution to Eq.\,(\ref{e:flow2}) 
with the initial condition Eq.\,(\ref{e:ini1}) can be formally given by the following form:
\aln{
\mathcal{L}^{(\alpha_0)}&=\mathcal{L}^{(0)}+\alpha_0\frac{d \mathcal{L}^{(\alpha_0)}}{d\alpha_0}\bigg|_{\alpha_0=0}+\frac{\alpha_0^2}{2}\frac{d^2 \mathcal{L}^{(\alpha_0)}}{d\alpha_0^2}\bigg|_{\alpha_0=0}+\cdots\nonumber\\
&=\mathcal{L}^{(0)} + \alpha_0 
F(s_{\mu\nu},M;g_{\mu\nu};\alpha_0)\Big|_{g_{\mu\nu}=\eta_{\mu\nu}}\,.
\label{e:fin1}
}
Here we have separated the undeformed Lagrangian and have explicitly extracted $\alpha_0$ 
as the overall factor of the deformation term. 

\medskip 

In the following, we write the trace and the traceless parts as $s$ and $\tilde{s}_{\mu\nu}$, respectively.
Now we can transform Eq.\,(\ref{e:fin1}) into a form with auxiliary fields, whose action is given by
\aln{
\frac{S}{N}&=\int\!\! d^2x\,\biggl[\mathcal{L}^{(0)}-\tilde{s}_{\mu\nu}\wt{C}^{\mu\nu}-MC-sB+H(\wt{C}^{\mu\nu},C,B;\alpha_0)  
\biggr]\,.
\label{e:fin2}
}
$H$ is determined so that Eq.\,(\ref{e:fin2}) should turn back to Eq.\,(\ref{e:fin1}) with the solutions of EoM for $\wt{C}^{\mu\nu}$, $C$ and $B$. This is the finite deformation counterpart of Eq.\,(\ref{e:act6}) 
\!\footnote{
In the infinitesimally deformed theory, we have focused on the lowest-order term in $F$. 
It does not include $s$ because it is written with the undeformed energy-momentum tensor:
\aln{
T^{(0)}_{\mu\nu}=2\tilde{s}_{\mu\nu}+Mg_{\mu\nu}\Big|_{g_{\mu\nu}=\eta_{\mu\nu}}\,. \notag
}
Thus we did not have to introduce $B$ as a source for $s$ in the infinitesimal case. 
}\,\,\,
(note that $\tilde{s}_{\mu\nu}\wt{C}^{\mu\nu}+MC=(1/2)T^{(0)}_{\mu\nu}C^{\mu\nu}$).

\medskip

There is one essential difference between Eqs.\,(\ref{e:fin2}) and (\ref{e:act6}); the introduction of a new scalar field $B$. In the following, we perform the large $N$ analysis on
Eq.\,(\ref{e:fin2}) just in the same manner as in the previous section. 

\medskip

Would it be convenient to interpret Eq.\,(\ref{e:fin2}) in terms of gravitational quantities, like Eq.\,(\ref{e:act7})? We have already parametrized all the components of $g_{\mu\nu}$ with $C_{\mu\nu}$. Therefore we must treat $B$ as an additional scalar field independent of the metric. In this case, however, we are not able to rewrite Eq.\,(\ref{e:fin2}) with $g_{\mu\nu}$. The action can be written down as follows: 
\aln{
\frac{S}{N}
=\!\!\int\!\! d^2x\,\sqrt{-g'}\,\qty[\frac{1}{2}g^{\prime\mu\nu}\partial_\mu\vec{\phi}\cdot\partial_\nu\vec{\phi}-\frac{m_0^2}{2}\vec{\phi}^{\,2}]
&+\int\!\! d^2x\,\sqrt{-g'}\,\Psi\qty[\frac{1}{2}g^{\prime\mu\nu}\partial_\mu\vec{\phi}\cdot\partial_\nu\vec{\phi}-\frac{m_0^2}{2}\vec{\phi}^{\,2}]
\nonumber\\
&-\int\!\! d^2x\,\frac{1}{2}B\,\eta^{\mu\nu}\partial_\mu\vec{\phi}\cdot\partial_\nu\vec{\phi}\,+\!\!\int\!\! d^2x\,H.
\label{e:fin3}
}
The new term involving $B$ explicitly violates the diffeomorphism invariance. 
Hence, it seems likely that there is no advantage even if we consider 
the action of the form (\ref{e:fin3}). 

\medskip

We would like to evaluate the kinetic terms for $\chi$ and $B$. 
The difference between this analysis and that in the previous sections is 
the presence of the mixing term of $\chi$ and $B$. Here, we can conclude 
the existence of the negative-norm mode by following the discussion around Eq.\,(\ref{e:mix1}) again. 
The induced kinetic terms are written as
\!\footnote{
The VEVs have been dropped off in this section. While its existence is guaranteed by the large $N$ limit, it does not play an essential role in proving the existence of the negative-norm mode. It is because the nonzero values of $\expval{C}$ simply rescale the flat metric, and $\expval{B}$ has nothing to do with the sign of $(\partial_\mu \chi)^2$.  
}\,\,\,
\aln{
\frac{\Gamma|_\text{quad}}{N}~\rightarrow~\frac{1}{2}\int \frac{d^2p}{(2\pi)^2} ~p^2\,\mqty(\chi(p)&B(p))\mqty(-\frac{1}{48\pi}&t\\t&u)\mqty(\chi(-p)\\B(-p)),
\label{e:ser2} 
}
where $t$ and $u$ are some constants. 
\!\!\!\footnote{
Note that in order to keep them finite, we need to add counterterms to the action in the form of $(\partial_\mu B)^2$ and $\partial_\mu B \partial^\mu\chi$. From the viewpoint of the original action (\ref{e:fin2}), these terms inevitably violate the locality.  
}\,\,\,
Thus we have at least one negative-norm mode.
%

\medskip

From the above analysis, we conclude that the general deformation of the theory with a function of the energy-momentum tensor has a negative-norm state even at the finite deformation. 
This is a non-perturbative result obtained by the large $N$ analysis. 

\medskip

On the other hand, there is no criterion so far in making Eq.\,(\ref{e:fin2}) covariant under the diffeomorphism. 
In order to realize the diffeomorphism invariance, it is critical to understand the gravitational meaning of $B$, 
which is coupled to other fields only at higher order in $\alpha_0$.

\section{Beyond the large $N$ limit?}
\label{s:1/N}
It is natural to ask how things would change when the $1/N$ corrections are taken into account. 
If the theory is modified to be diffeomorphism-invariant, the $1/N$ corrections can be evaluated rather easily.

\medskip

\subsection*{In the ordinary gravity}

First, let us make a short review of the ordinary 2d gravity, which would be 
the simplest example of diffeomorphism-invariant theories \cite{Knizhnik:1988ak, Distler:1988jt, David:1988hj}. In a 2d CFT coupled to gravity, it is standard to take the conformal gauge as 
$g_{\mu\nu}=e^{\varphi}\hat{g}_{\mu\nu}$ and to integrate the conformal mode $\varphi$. 
One can convert $\varphi$ to an ordinary scalar field defined on the background metric $\hat{g}_{\mu\nu}$ by taking account of the path-integral measure as follows.  
The original measure $\mathcal{D}_g\varphi$ is defined by the following metric:
\aln{
|\!|\delta\varphi|\!|_g^2\equiv\int d^2x\sqrt{-g}(\delta\varphi(x))^2=\int d^2x\sqrt{-\hat{g}}e^{\varphi}(\delta\varphi(x))^2.
}
On the other hand, the ordinary scalar field on the background metric $\hat{g}_{\mu\nu}$ has the measure $\mathcal{D}_{\hat{g}}\varphi$ defined by the following metric: 
\aln{
|\!|\delta\varphi|\!|_{\hat{g}}^2\equiv\int d^2x\sqrt{-\hat{g}}(\delta\varphi(x))^2.
}
One can show that these two measures, $\mathcal{D}_g\varphi$ and $\mathcal{D}_{\hat{g}}\varphi$, are related as
\aln{
\mathcal{D}_g\varphi=\mathcal{D}_{\hat{g}}\varphi\,\exp\left({\frac{1}{48\pi}S_\text{L}\bigg|_{g_{\mu\nu}=e^\varphi\hat{g}_{\mu\nu}}}\right). 
}
Therefore the total effective action for $\varphi$ takes the following form:
\aln{
\Gamma=\frac{25-N}{48\pi}\,S_\text{L}\bigg|_{g_{\mu\nu}=e^\varphi{\hat{g}}_{\mu\nu}}+\int d^2x\sqrt{-\hat{g}}\,\Lambda:e^{\alpha'\varphi}\!:_{\,\hat{g}},
\label{e:SL4}
}
where we have renormalized the cosmological constant term. 
The notation $:e^{\alpha'\varphi}\!:_{\,\hat{g}}$ denotes the normal-ordered operator for $e^{\alpha'\varphi}$ and 
a constant $\alpha'$ is determined by the fact that Eq.\,(\ref{e:SL4}) should be invariant 
under the transformation $\hat{g}_{\mu\nu}\to e^\sigma\hat{g}_{\mu\nu}\,,~\varphi\to\varphi-\sigma$. The explicit value of $\alpha'$ is given by
\aln{
\alpha'=\frac{25-N-\sqrt{(25-N)(1-N)}}{12}.
\label{e:alpha'}
}
The system with the cosmological constant is positive-definite only when $N\leq1$. 
(On the other hand, when the renormalized cosmological constant is tuned to be zero, 
the quantum fluctuation of $\varphi$ can be integrated out as long as $N<25$.) 
The crucial point is that the path-integral of the fluctuation of $\varphi$ amounts only to the shift of the kinetic term in Eq.\,(\ref{e:SL4}), and the renormalization of the cosmological constant (\ref{e:alpha'}). $\varphi$ is now a free field on the background $\hat{g}_{\mu\nu}$, in that the interactions generated by the cosmological constant term do not modify the kinetic term by radiative corrections. 
This conclusion also holds in a general field theory as long as it is approximated by a CFT 
in the high-energy limit, by regarding $N$ as the central charge of the CFT.

\medskip

The crucial observation here is that the quantum effect of the conformal mode amounts 
just to the shift of $N$ by unity. In other words, the $1/N$ correction appears 
in the effective action as a quantity of $1/N$ without a large numerical factor.

\subsection*{In our theory}

Let us go back to our theory Eq.\,(\ref{e:act8}). In this case, $\chi$ corresponds to the conformal mode $\varphi$ 
in the above discussion as $\varphi=\log(1+\chi)$, as can be seen from Eq.\,(\ref{e:metrep}). 
However, Eq.\,(\ref{e:act8}) does not have the diffeomorphism invariance because $h$ and $\Psi$ are not invariant. 
They produce the extra contributions to the effective action for $C^{\mu\nu}$. In particular, they induce corrections in the kinetic terms. 

However, 
we still expect that the $1/N$ corrections to Eq.\,(\ref{e:act8}) are of the same order as that in the ordinary gravitational theory. As a result, the conclusion in the previous section should hold even when 
the $1/N$ corrections are taken into account.  
If it is true, the emergence of the negative norm state is not avoidable unless the central charge $N$ is of $O(1)$.  

%
%
%
%

\section{Summary}
\label{s:sum}

In this paper, we have performed the large $N$ analysis to study quantum aspects of the $O(N)$ vector model that is deformed by a general function of $T_{\mu\nu}$, including the  $T\bar{T}$-deformation. Bound states of the original field appear and are of negative norm. 
The negative-norm mode can be understood by comparing the theory with the one coupled to gravity. 
It corresponds to the conformal mode of the metric, whish is described by the Liouville action. As for the two-point function in the large $N$ limit, 
there is no difference between our theory and the ordinary gravity. Hence, in conclusion, the theory non-perturbatively violates positivity at least in the large $N$ limit, whether the deformation is infinitesimal or finite. 

\medskip

In order to remedy it, some degrees of freedom with negative central charge need to be introduced. The natural candidates are the FP ghosts, and accordingly, diffeomorphism invariance should be equipped. Then the theory becomes positive-definite for $N<25$.  
Without the diffeomorphism invariance, we expect that the numerical factor of the $1/N$ correction should not be too large. Unless $N$ is of $O(1)$, the negative-norm states will unavoidably emerge. To study this correction remains as a future problem. 

\medskip

Another important point is that the finite deformation (namely, higher order corrections in $\alpha_0$) induces $B$ in addition to the ordinary gravity. While $B$ might be interpreted as a kind of dilaton, the diffeomorphism invariance has not been manifest so far. To investigate this issue is 
an interesting open question.

\section*{Acknowledgements}
The work of H.K.\ was supported in part by JSPS Grant-in-Aid for Scientific Research (C) No.\,16K05322 .
The work of T.I.\ was supported in part by JSPS Grant-in-Aid for Scientific Research (C) No.\,19K03871.
The work of K.S.\ was supported by Grant-in-Aid for JSPS Research Fellow Number 17J02185. 
The works of T.I.\ and K.Y.\ were supported by the Supporting Program for Interaction-based Initiative 
Team Studies (SPIRITS) from Kyoto University, and JSPS Grant-in-Aid for Scientific Research (B) 
No.\,18H01214. This work is also supported in part by the JSPS Japan-Russia Research Cooperative Program.  


\vspace*{2cm} 

\section*{Appendix}

\appendix
\section{Derivation of the induced kinetic term}
\label{s:cal}

This Appendix is devoted to explaining the details of the derivation of the induced kinetic term. 

\medskip 

First of all,  we will evaluate the trace term in Eq.\,(\ref{e:setwo}) explicitly as  
\begin{align}
&\tr\qty[\frac{1}{-\partial^2-m^2-i\vae}(-m_0^2C-\partial_{\mu}\wt{C}^{\mu\nu}\partial_{\nu})]^2
\nonumber\\
&=\int d^2x\,d^2y\,d^2z\,d^2w\biggl\{
\bra{x} \frac{1}{-\partial^2-m^2+i\vae}
\ket{y}\bra{y}\qty(m^2C+\partial_{\rho}\wt{C}^{\rho\sigma}\partial_{\sigma})
\ket{z}\nonumber\\
&~~~~~~~~~~~~~~~~~~~~~~~~~~~~~~~\times\bra{z}\frac{1}{-\partial^2-m^2+i\vae}
\ket{w}
\bra{w}\qty(m^2C+\partial_{\mu}\wt{C}^{\mu\nu}\partial_{\nu})
\ket{x}\biggr\} \nonumber 
\\
&=\int d^2x\,d^2y\biggl\{
C(x)\qty(m_0^4G(x-y)^2)C(y)\nonumber\\
&~~~~~~~~~~~~~~~~~~~~~~~~~~~~-2C(x)\qty(m_0^2\pdv{y^{\mu}}G(x-y)\pdv{y^{\nu}}G(y-x))\wt{C}^{\mu\nu}(y)
\nonumber\\
&~~~~~~~~~~~~~~~~~~~~~~~~~~~~+\wt{C}^{\mu\nu}(x)\qty(\pdv{}{x^{\mu}}{y^{\rho}}G(x-y)\pdv{}{x^{\nu}}{y^{\sigma}}G(y-x))\wt{C}^{\rho\sigma}(y)\biggr\},
\end{align}
where the propagator $G(x-y)$ is as usual   
\begin{align}
G(x-y)\equiv\bra{x}\frac{1}{-\partial^2-m^2+i\vae}\ket{y}=\int\!\! \frac{d^2k}{(2\pi)^2}\frac{e^{ik\cdot (x-y)}}{k^2-m^2+i\vae}\,.
\end{align}
Thus the effective action takes the following form:  
\begin{align}
\label{e:se}
\Gamma&=\frac{1}{2}\int dx \int dy \mqty(C(x)&\wt{C}^{\mu\nu}(x))
\mqty(M&M_{\rho\sigma}\\M_{\mu\nu}&M_{\mu\nu\rho\sigma})
\mqty(C(y)\\\wt{C}^{\rho\sigma}(y))\,,
\\M&\equiv\frac{m_0^4}{2i}\qty[G(x-y)]^2-\frac{1}{2\alpha_0}\delta^2(x-y)\,,
\\M_{\mu\nu}&\equiv-\frac{m_0^2}{2i}\pdv{x^{\mu}}G(x-y)\pdv{x^{\nu}}G(y-x),
\\M_{\mu\nu\rho\sigma}&\equiv\frac{1}{2i}\pdv{}{x^{\mu}}{y^{\rho}}G(x-y)\pdv{}{x^{\nu}}{y^{\sigma}}G(y-x) \notag \\
& \qquad +\qty(\frac{1}{4\alpha_0}+(1+\expval{C})\Lambda_0)\eta_{\mu\rho}\eta_{\nu\sigma}\delta^2(x-y)\,.
\end{align}
Practically, $M_{\mu\nu}$ and $M_{\mu\nu\rho\sigma}$ are projected out so that they should be symmetric traceless with respect to indices $(\mu\nu)$ and $(\rho\sigma)$, because they are contracted with $\wt{C}^{\mu\nu}$. Each term corresponds to one of the diagrams depicted in Fig.\,\ref{f:prop}.

\begin{figure}[t]
\centering
\subfigure[$M$]
{\includegraphics[scale=0.3]{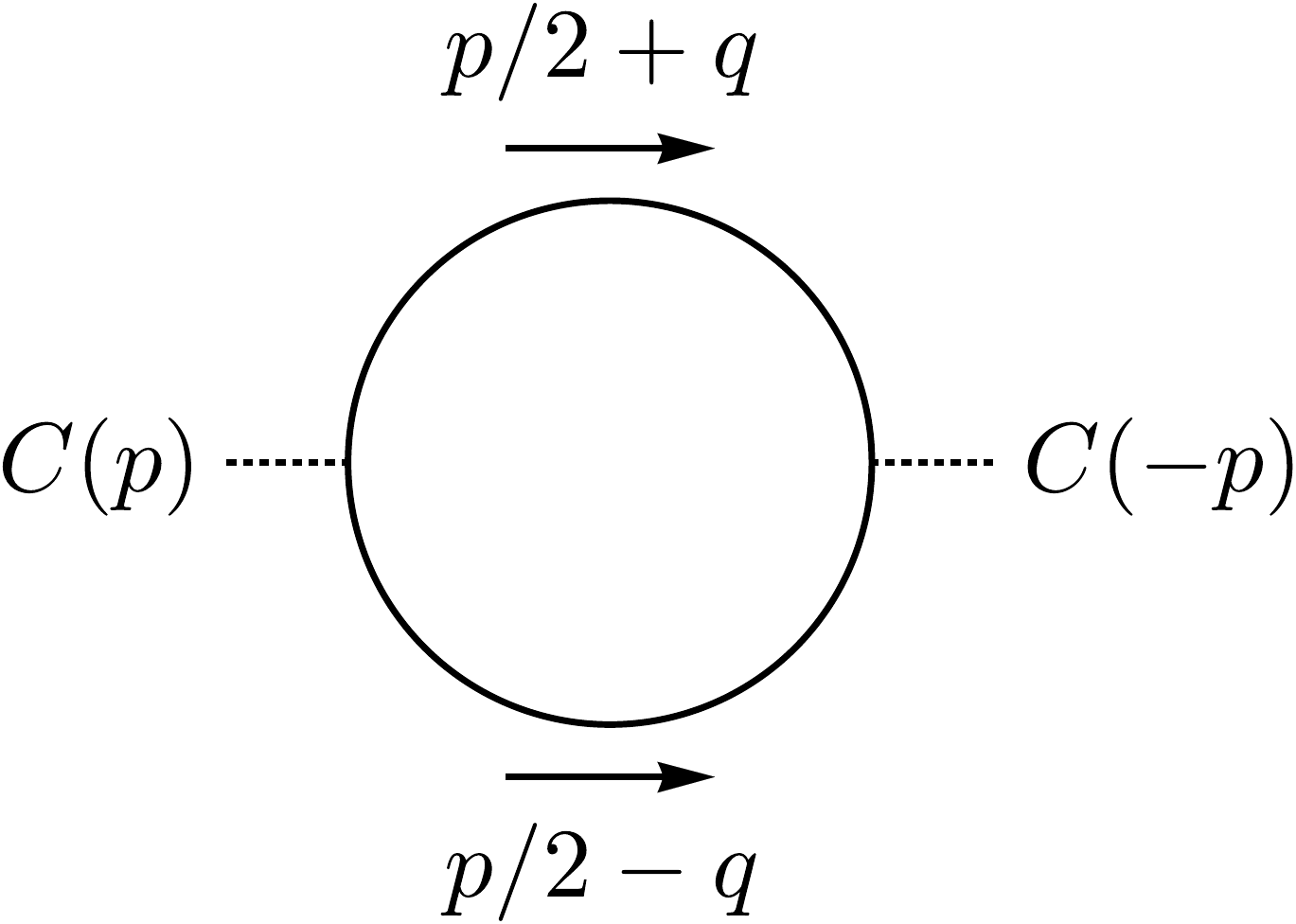}
\label{f:loopM}}
\subfigure[ $M_{\mu\nu}$]{
\includegraphics[scale=0.3]{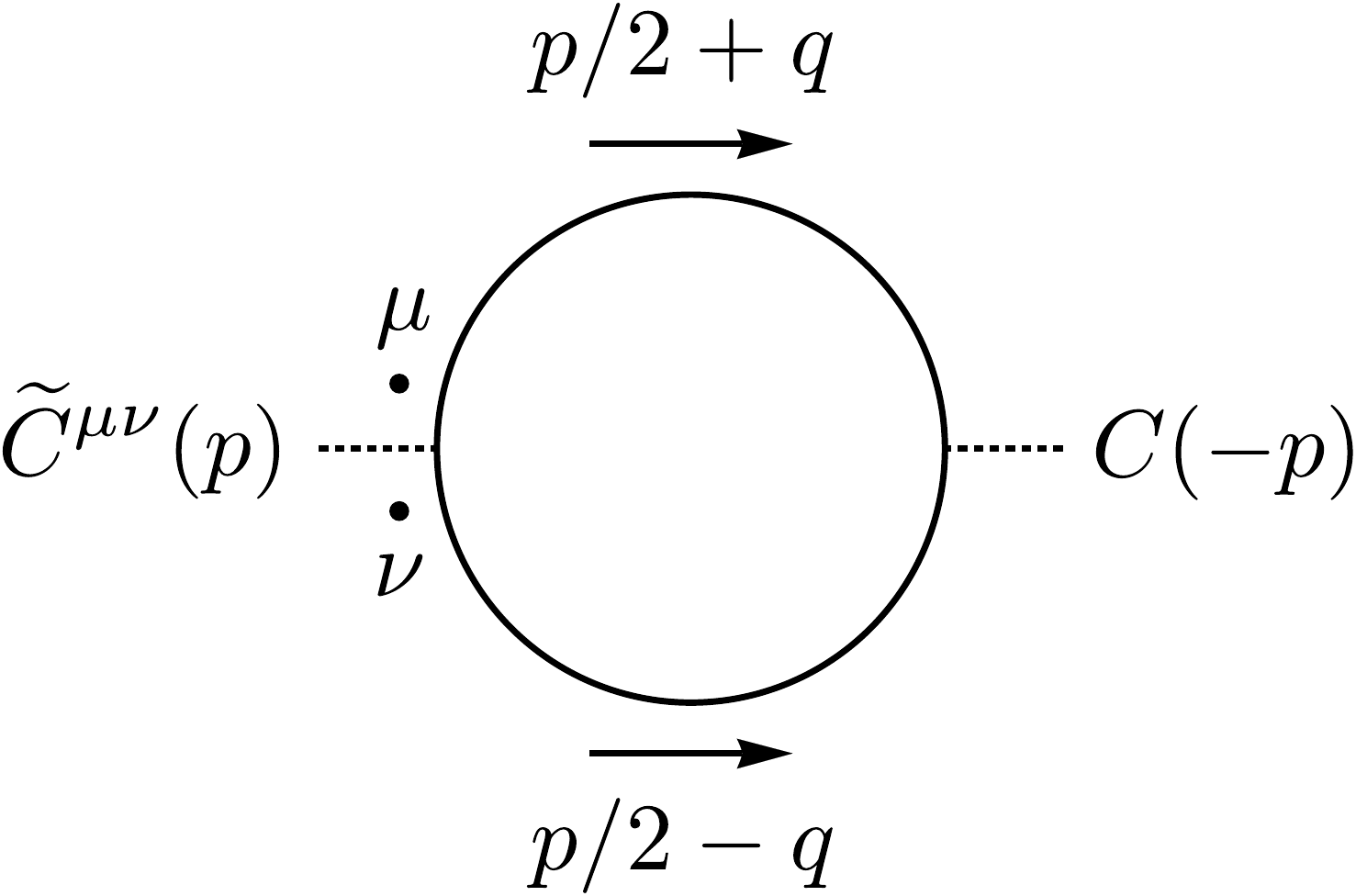}
\label{f:loopMmunu}}
\subfigure[ $M_{\mu\nu\rho\sigma}$]{
\includegraphics[scale=0.3]{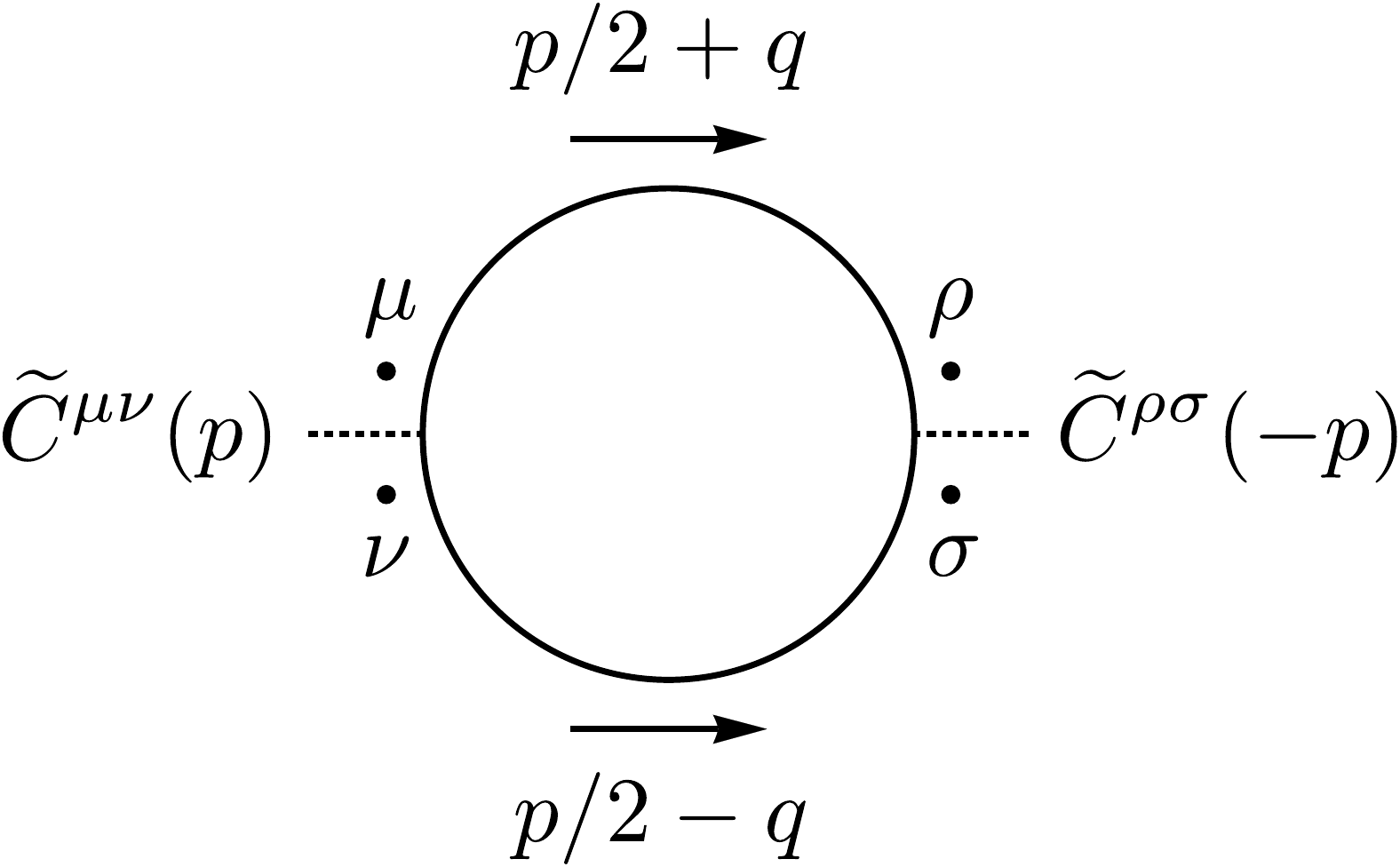}
\label{f:loopMmnrs}}
\caption{Feynman diagrams corresponding to $M$\,, $M_{\mu\nu}$ and $M_{\mu\nu\rho\sigma}$\,.}
\label{f:prop}
\end{figure}

After moving to the momentum space, it is easy to perform the loop integrals with PV regularization.
The results are listed below:
\footnote{
The normalization of functions in the momentum space here is taken as
\aln{
\int\! d^2x\,f(x)=\int\!\frac{d^2p}{(2\pi)^2}\,f(p).
}
}
\begin{align}
&M(p)
=g(p)-\frac{1}{2\alpha_0},
\label{e:Mdef}
\\
&M_{\mu\nu}(p)
=-\frac{p_\mu p_\nu}{p^2}g(p),
\label{e:Mmunudef}
\\
&M_{\mu\nu\rho\sigma}(p)=
 \frac{p_{\mu}p_{\nu}p_{\rho}p_{\sigma}}{p^4}g(p)
+\eta_{\mu\rho}\frac{p_{\nu}p_{\sigma}}{p^2}h(p)
+\eta_{\mu\rho}\eta_{\nu\sigma}f(p)\nonumber\\
&~~~~~~~~~~~~~~~~~~~~~~~~~~~~~~~~~~~~~~~+\qty(\frac{1}{4\alpha_0}+(1+\expval{C})\beta_0)\eta_{\mu\rho}\eta_{\nu\sigma},
\label{e:M4def}
\end{align}
where scalar functions $f(p)$\,, $g(p)$ and $h(p)$ are defined as 
\begin{align}
f(p)& \equiv -\frac{h(p)}{2}
+\frac{1+\expval{C}}{16\pi} \left[ \qty(M_1^2-m_0^2)\frac{r}{1-r}\log(r)+\log(\frac{M_1^2}{m_0^2}) \right],
\\g(p)& \equiv -\frac{m_0^2}{8\pi }-\frac{p^2}{48 \pi }+\frac{m_0^4 }{2\pi  p \sqrt{4 m^2-p^2}}\tan ^{-1}\left(\frac{p}{\sqrt{4 m^2-p^2}}\right),
\\h(p)& \equiv 
\frac{ m_0^2}{6\pi}+\frac{p^2}{144\pi}\left[ 3\log \left(\frac{m_0^2}{M_1^2}\right)+\frac{3 \log (r)}{r-1}-8\right]\nonumber\\
&~~~~~~~~~~~~~-\frac{ \left(4 m_0^2-p^2\right)^{\frac{3}{2}} }{24\pi p}\tan ^{-1}\left(\frac{p}{\sqrt{4 m^2-p^2}}\right).
\end{align}
An important point is that both of the finite and divergent parts in the coefficient of $p^2(\wt{C}^{\mu\nu})^2$ completely vanishes. One can easily check this by using the following identity: 
\aln{
\int\!\frac{d^2p}{(2\pi)^2}\,F(p^2)\wt{C}_\mu^{~\lambda}(p)\wt{C}_\lambda^{~\nu}(-p)=\int\!\frac{d^2p}{(2\pi)^2}\,F(p^2)\frac{1}{2}\delta_\mu^{~\nu}\left(\wt{C}_\lambda^{~\rho}(p)\wt{C}_\rho^{~\lambda}(-p)\right),
}
where $F(p^2)$ is an arbitrary function of $p^2$. The reason for the absence of $p^2(\wt{C}^{\mu\nu})^2$ term is explained in Section \ref{s:grav}. On the other hand, the mass terms for $C'$ and $\wt{C}^{\mu\nu}$ 
are given by 
\aln{
\int\!\! \frac{d^2p}{(2\pi)^2}&\biggl[-\frac{1}{4\alpha_0}C'(p)C'(-p)\nonumber\\
&+\biggl\{\frac{1}{8\alpha_0}+\frac{1+\expval{C}}{2}\biggl(\beta_0-\frac{1}{32\pi}\biggl[m_0^2\log(\frac{m_0^2}{M_1^2})-\frac{r\log(r)}{1-r}M_1^2\nonumber\\
&~~~~~~~~~~~~~~~~~~~~~~~~~~~~~~~~~~~~~~~~~~~~~~~~~~~~~~~~~~~~+\frac{r\log(r)}{1-r}m_0^2\biggr]\biggr)\biggr\}\wt{C}^{\mu\nu}(p)\wt{C}_{\mu\nu}(-p)\biggr].
}
The divergent contribution should be absorbed into $\beta_0$\,. By noting that it is just the minus quarter 
of the divergent correction to $\Lambda_0$, the parameter $\beta_0$ can be set as $\beta_0=-\Lambda_0/4$.


\end{document}